\documentstyle[preprint,aps]{revtex}

\begin{document}
\title{ The Klein-Gordon oscillator and the proper time formalism 
in a Rigged Hilbert sapace}
\author{A. G. Grunfeld \thanks{Electronic address: 
grunfeld@venus.fisica.unlp.edu.ar} and M. C. Rocca}
\address {Departamento de F\'{\i}sica, Fac. de Cs. Exactas,
Universidad Nacional de La Plata.\\
C.C. 67 (1900) La Plata, Argentina.}
\maketitle

\begin{abstract}

The implications of manifestly covariant formulation of relativistic
quantum mechanics depending on a scalar evolution parameter, canonically
conjugated to the variable mass, is still an unsettled issue. In this
work we find a complete set of generalized eigenfunctions of the Klein-Gordon 
Oscillator in the above mentioned formulation, in the Rigged 
Hilbert Space with Tempered Ultradistributions. We briefly comment on
some models where this solution could be applied.


\end{abstract}

\vspace{1cm}
\begin{center}
{\it To appear in Il Nuovo Cimento B}
\end{center}
\newpage

\section{Introduction}

One of the main features of the relativistic mechanics is that
time and space coordinates are on the same footing. In contrast,
in standard relativistic quantum mechanics 
the treatment of those quantities is not symmetric.
This fact poses a puzzle to physics since long time ago.       
In a fine set of papers, Dirac introduced the idea of
promoting the time coordinate $x^0$ to the rank of operator \cite {dirac}. 
This proper time formalism introduces an absolute evolution parameter 
(related to the proper time in the classical limit), 
which parametrizes the dynamics of the quantum system. 
The basic idea of these works have probably fallen into
oblivion because Dirac himself 
did not insist on it in his celebrated paper of 1928 \cite{Dirac}. 
This formalism shortly returned to view with Feynman and St\"{u}ckelberg 
antiparticles interpretation as particles moving backward in time 
\cite{Fey,Stu}.
A lot of progress have been done in oder to enlight this ideas
\cite{Ap,CF,Kalo,Phil,Ky}. 
We think that this formalism is the natural generalization of the 
relativistic quantum mechanics which gives a consistent 
way to study quantum systems.
In this scenario, we work out the harmonic oscillator (HO) solution of the 
Klein--Gordon (KG) equation in the Rigged Hilbert Space (RHS) with Tempered
Ultradistributions (for an explanation about Rigged Hilbert Spaces
and physical applications see ref.\cite{Bohm}) 
adopting the Collins and Fanchi (CF) formalism \cite{CF}.
In ref.\cite{bcdr} it is shown how to work out the quantum mechanics
in an abstract RHS. The results obtained in our paper
are derived from those obtained in ref.\cite{bcdr} and \cite{mario}. 
As Feynman pointed out \cite{Fey}, the proper time formalism
introduces exponentially increasing solutions in the motion equations.
The introduction of the RHS $(h, {\cal{H}}, {\Lambda}_{\infty})$ 
and in $(H, {\cal{H}}, {\cal{U}} )$ has allowed us to solve this problem. 
With this tools, our work contributes 
enrich the study of relativistic quantum systems in terms of the 
proper time theory.

\indent We have also a more practical motivation. Our solutions could be
also relevant because the HO is present in most problems of physics
describing different interactions and we use a consistent mechanism
to obtain the eigenvalues and eigenfunctions. 
In the last decades many works have been done in order to
describe, under a quantum relativistic frame, the harmonic
oscillator. 
Kim and Noz \cite{Kim} proposed a covariant HO in a hyperplane
formalism to introduce it in the quark model. They had also studied
excited meson decays with this technique. Moshinsky et al have
introduced a new kind of interaction in Dirac equation which in
the non-relativistic limit becomes an harmonic oscillator with a
very strong spin-orbit coupling term. Later on, this procedure was
followed by Moreno and Zentella \cite{Moreno} to take into account 
the quark-antiquark interaction giving the mass spectra for quarkonium
systems.

In our work, we present the eigenvalues and eigenfunctions for the harmonic
oscillator in the proper time formalism. The solutions we have
obtained contains, as a particular case, those for the usual
KGHO. Furthermore, we have solutions for particles
with negative mass.

The paper is organized ls follows: in Sect. 2 we present the proper
time formalism. Sect. 3 is devoted to the study of the harmonic
oscillator in the above mentioned frame. In Sect. 4 we discuss
the results.

\section{The Proper-time formalism}

Let us briefly sum up the characteristic of the formalism.
The square of the amplitude of the wave function $\Psi$ must be interpreted 
as the probability of an event corresponding to the 
position of a particle at a particular time. It is worthwhile to
point out that the latter interpretation differs from the usual one due to 
that there is a probability distribution associated with time.
Therefore, for each $t$ there is a probability  that a particle may
or may not be found somewhere in space (as in particle decays).  
Then, the mean value of the position operator $\hat{x}$, can be written as 
\begin{equation}
< x^{\mu} > = \int_{ST} x^{\mu} \rho \; d ^4 x
\end{equation}   
where $\rho$ is a distribution on the spacetime manifold satisfying the
two following conditions: 
\begin{displaymath}
\rho > 0 \;\;\;\;\;\;\;\;\;\;{\rm and} \;\;\;\;\;\;\;\;\;\; \int_{ST}
\rho \; d ^4 x = 1
\end{displaymath}
with integration understood in the Lebesgue sense, over the
entire spacetime manifold.

Taking this into account, the on-shell condition  
must be understood as the expectation value of the $p^{\mu} p_{\mu}$ 
operator. As shown by CF, the aforementioned operator 
is proportional to $\partial / \partial\tau$, and the light-cone 
constraint identifies 
$\tau$ with the classical proper time \cite{CF}. Thereby, the free particle 
equation, in natural units, is given by
\begin{equation}
i \;2\; \bar{m} \frac{\partial \Psi}{\partial\tau} = H \Psi 
\label{ga}
\end{equation}
here $H = p^{\mu}p_{\mu}$ and $\bar{m}$ is defined as the classical limit 
of the $p^{\mu}p_{\mu}$ expectation value.

Our purpose is to work out the relativistic spinless harmonic oscillator
in the frame of the formalism we have commented.

\section{Klein-Gordon Oscillator}
The hamiltonian for the relativistic harmonic oscillator was studied by 
Moshinsky and Szczepaniak \cite{BM}. They proposed a new type of interaction
in the Dirac equation, linear in coordinates and momentum. The corresponding
equation has been named ``Dirac Oscillator'' because in the non-relativistic 
limit the harmonic oscillator has been obtained. This kind of interaction was 
introduced in the Klein-Gordon equation \cite{Bruce,Vale}.
To get a closed form for the extended relativistic harmonic oscillator 
hamiltonian, we present first the Klein-Gordon oscillator developed in 
the Bruce \cite{Bruce} formalism together 
with the Sakata-Taketani approach \cite{Saka}. The latter selection 
determines the following KG equation 
\begin{equation}
\left( \Box + \, m^2 \, \omega^2 \, r^2 \, 
- \, 3 \, m \, \omega \, + \, m^2 \right)  \Psi \; = \;0  
\label{001}
\end{equation}

\section{Scalar-time parametrization of Klein-Gordon Oscillator}

Now, we want to work out this equation in the frame of the scalar time 
parametrization
taking into account the above definition of the $p^{\mu}p_{\mu}$ operator
and considering a probability distribution asociated with time. 
To solve this equation we shall consider the wave function
$\Psi(x_{\mu}, \tau)$  as an
element of the space of exponentially increasing distributions 
(in the variable $\tau$)
$\Lambda_{\infty}$ \cite{tp9}. The space $\Lambda_{\infty}$ is formed
by distributions T of the exponential type satisfying:
\begin{equation}
T = \left( {\partial}^k / \partial x^k  \right)  \left[ e^{k|x|} f(x) \right]
\end{equation}
where k is an integer greater than or equal to zero and $f$ is
bounded continuous.\\
$\Lambda_{\infty}$ is the dual of the space H of all functions
$ \phi \in C^{\infty}$ in R (real numbers) such that 
$ e^{k|x|} D^p \phi (x) $ is bounded in R for all k and p . \\
If $\cal{H} $ is the Hilbert space of square integrable functions,
the triplet $(H, {\cal{H}}, {\Lambda}_{\infty})$ is a Rigged Hilbert 
Space (RHS) \cite{bcdr}. The Fourier transformed triplet of
$(H, {\cal{H}}, {\Lambda}_{\infty})$ is the RHS
$(h, \cal{H}, \cal{U})$. In this triplet $h$ is the space of analytical
test functions, rapidly decreasing in any horizontal band. We denote by
$a_{\omega}$ the space of all functions $F(z)$ such that: \\
i) $F(z)$ is analytic in $\{z \in C : |Im(z)| > k \}$ \\
ii) $ F(z) / z^k$ is bounded continuous in 
$\{z \in C : |Im(z)| \geq k \}$ where k is an integer depending on $F(z)$.
Let $\Pi$ be the set of all polynomials in the variable z. It has been
demonstrated in ref.\cite{tp9} that
${\cal{U}} = a_{\omega} / \Pi $ where $\cal{U}$ is by definition the space of
Tempered Ultradistributions \cite{tp9,tp8,tp11,tp10}. 
In the RHS $(H, {\cal{H}}, {\Lambda}_{\infty})$ 
(and in $(h, {\cal{H}}, {\cal{U}} ) $ ) a linear
and symmetric operator $A$ acting on $H (h)$ which admits a self-adjoint
prolongation $\bar{A}$ acting on $\cal{H}$, has a complete set of
eigen-functionals on ${\Lambda}_{\infty} ({\cal{U}})$ with real generalized 
eigenvalues \cite{GE1,GE2}. \\

According to ref.\cite{GE3}, if $ f \in {\Lambda}_{\infty} $,
$ \phi \in H $ , $ \hat{f} \in {\cal{U}} $ and $ \hat{\phi} \in h $
we have:
\[<f,\phi > = \int\limits_{-\infty}^{+\infty} \overline{f(\tau)} 
\phi (\tau) \; d\tau =\]
\begin{equation}
\int\limits_{\Gamma} \hat{f}({\alpha}) \hat{\phi} (\alpha) \;
d\alpha = <\hat{f}, \hat{\phi} >
\end{equation}
where $ \Gamma $ is the path which surrounds all the singularities
of $ \hat{f} (\alpha) $ placed on a 2k wide band that encircles the
real axis. The path $ \Gamma $ runs from $ -\infty $ to $ + \infty $
along $ Im(\alpha) > k $ and from $ +\infty $ to $ -\infty $ along
$ Im(\alpha) < -k $ ( ref.\cite{tp9} ) and
\begin{equation}
\overline{f(\tau)} = \int\limits_{\Gamma} \hat{f} (\alpha)
e^{-i \tau \alpha}\; d\alpha = {\cal{F}}^{-1} \{\hat{f}(\alpha) \}
\end{equation}
\begin{equation}
\phi(\tau) = \frac {1} {2\pi} \int\limits_{-\infty}^{+\infty}
\hat{\phi}(\alpha) e^{i \tau \alpha} \; d\alpha = {\cal{F}}^{-1}
\{\hat{\phi}(\alpha) \}
\end{equation}

We can define
$(2\; \bar{m}\; i \frac{\partial}{\partial\tau})^{1/2}$ operating over a 
$\Lambda_{\infty}$ distribution as follows \cite{mario}:  
\begin{equation}
\left(2\; \bar{m}\; i \frac{\partial}{\partial\tau}\right)^{1/2} f(\tau) =
{\cal{F}}^{-1}  \left\{ \alpha^{1/2} \left( \hat{f}(\alpha) + a(\alpha) \right) \right\}
\end{equation}
where
\begin{equation}
\hat{f}(\alpha) = {\cal{F}} \{ f(\tau) \}
\end{equation}
and $ a(\alpha) $ is an entire analytical function.
In view of this definition and (5), (6) and (7) it follows that the
operator $(i \partial / \partial \tau)^{1/2} $ is linear and self-adjoint.

With all these requirements eq. (\ref{001}), takes the form:
\begin{equation}
\left[ \Box + 2 \, \overline{m} \, i \, \omega^2 \, r^2 \, 
\frac{\partial}{\partial \tau} \, - \, 3 \, \omega \, 
\left( 2 \, \overline{m} \, i \, \frac{\partial}{\partial \tau} \right)^{1/2} \,+ 
\,2 \,\overline{m} \, i \frac{\partial}{\partial \tau} \right]  \Psi \; = \;0  
\label{007}
\end{equation}
Thus, taking into account the Fourier transform,
equation (\ref{007}) could be written as follows :
\begin{equation}
\int_{\Gamma}\left( \Box + \omega^2 r^2 \alpha - 3 \omega \alpha^{1/2} + \right.
\left. \alpha \right) \left[ \hat{\Psi_{c}}(x_{\mu}, \alpha) + a (x_{\mu}, \alpha) \right ]
e^{-{\frac{i \alpha}{2 \overline{m}}\tau}} d\alpha = 0
\label{100}
\end{equation}
where $\Gamma$ is the usual path which surrounds all the singularities of
$\hat{\Psi_c}$ and $a(x_{\mu}, \alpha)$ is an entire analytical function
of the variable $\alpha$.
Defining the operator 
\begin{equation}
\Box + \omega^2 r^2 \alpha - 3 \omega \alpha^{1/2} + \alpha = L
\end{equation}
then, equation (\ref{100}) is equivalent to
\begin{equation}
L \left[ \hat{\Psi_c}(x_{\mu}, \alpha) + a(x_{\mu}, \alpha) \right] = 0
\end{equation}
Thus, we can write:
\begin{equation}
\hat{\Psi_c}(x_{\mu}, \alpha) = \hat{f_c}(x_{\mu}, \alpha) - a(x_{\mu}, \alpha)  
\end{equation}
with $\hat{f_c}$ a general solution of the homogeneus system,
\begin{equation}
L \hat{f_c}(x_{\mu}, \alpha) = 0
\end{equation}

The boundary condition we shall impose is to set the entire analytical
function ($a(x_{\mu},\alpha)$) equal to zero; with this condition we
ensure that for a fixed $\alpha > 0$ we obtain the same energy spectrum
as in the usual KG formalism. Therefore we obtain, 
\begin{equation}
\int_{\Gamma}\left( \Box + \omega^2 r^2 \alpha - 3 \omega \alpha^{1/2} + 
\alpha \right) \hat{f_c}(x_{\mu}, \alpha) 
e^{-{\frac{i \alpha}{2 \overline{m}}\tau}} d\alpha = 0
\end{equation}
On the real axis this expression takes the form,
\begin{eqnarray}
\int_{-\infty}^{+\infty} \left[ \left( \Box + \omega^2 r^2 \alpha - 
3 \omega (\alpha + i0)^{1/2} + \alpha \right) \hat{f_c}(x_{\mu}, 
\alpha + i0) -  
\right.& & \nonumber \\
\left. \left( \Box + \omega^2 r^2 \alpha -      
3 \omega (\alpha - i0)^{1/2} + \alpha \right) \hat{f_c}(x_{\mu}, 
\alpha - i0) \right]  
e^{-{\frac{i \alpha}{2 \overline{m}}\tau}} d\alpha & = & 0
\label{101}
\end{eqnarray}
If we impose to $\hat{f}_c$ to satisfy the equality 
\begin{equation}
\int_{-\infty}^{0} \left[ (\alpha + i0)^{1/2} \hat{f_c}(x_{\mu}, 
\alpha + i0) -  
(\alpha - i0)^{1/2} \hat{f_c}(x_{\mu}, \alpha - i0) \right]  
e^{-{\frac{i \alpha}{2 \overline{m}}\tau}} d\alpha = 0
\end{equation}
equation (\ref{101}) reads,

\[\int_{0}^{+\infty}  \left( \Box + \omega^2 r^2 \alpha - 
3 \omega \alpha^{1/2} + \alpha \right) \left[ \hat{f_c}(x_{\mu}, 
\alpha + i0) -  
\hat{f_c}(x_{\mu}, \alpha - i0) \right]  
e^{-{\frac{i \alpha}{2 \overline{m}}\tau}} d\alpha +  \]
\begin{equation}
\int_{- \infty}^{0} \left( \Box + \omega^2 r^2 \alpha  
+ \alpha \right) \left[ \hat{f_c}(x_{\mu}, \alpha + i0) -  
\hat{f_c}(x_{\mu}, \alpha - i0) \right]  
e^{-{\frac{i \alpha}{2 \overline{m}}\tau}} d\alpha = 0 
\end{equation}
Next, we define as in ref. \cite{tp8,tp9}
\begin{equation}
\hat{f} (x_{\mu}, \alpha) =  \hat{f_c} (x_{\mu}, \alpha + i0) - 
\hat{f_c} (x_{\mu}, \alpha - i0)
\end{equation}
Bearing this in mind, our system can be reduced to
\begin{equation}
\left( \Box + \omega^2 r^2 \alpha - 
3 \omega \alpha^{1/2} + \alpha \right) \hat{f}(x_{\mu}, \alpha ) = 0 
\;\;\;\;\alpha > 0  \\
\label{15}
\end{equation}

\begin{equation}
\left( \Box + \omega^2 r^2 \alpha +  \alpha \right)\hat{f}(x_{\mu}, \alpha ) = 0 \;\;\;\;\alpha < 0 
\label{16}
\end{equation}
which is solved using the standard procedure of 
quantum mechanics. Thus, introducing 
$\hat{f} \, = \, e^{-\,i \,E\, t} \,\; \varphi$, eq. (\ref{15}) becomes
\begin{equation}
(- \, \bigtriangleup + \,\alpha \, \omega^2 \, r^2 \, 
 - \, 3 \,\alpha^{1/2}\, \omega + \alpha \,) \,\; \varphi \; 
 = \;E^2 \,\varphi  
\label{010}
\end{equation}

The eingenfunctions that satisfy the last equation are:
\begin{eqnarray}
\varphi_{_{\alpha\;\,N_1\,N_2\,N_3}} \; & = & \;\, 
A_{_{\alpha\;\,N_1\,N_2\,N_3}} 
e^{- \alpha^{1/2} (\,x^2 \,+\,y^2\,+\,z^2\,)\;\, \omega^2 \,/\,2 } 
\nonumber \\
 & \times & \;\, H_{_{N_1}}(\sqrt{\alpha^{1/2} \omega}\,x)
\;H_{_{N_2}}(\sqrt{\alpha^{1/2} \omega}\,y)\;H_{_{N_3}}
(\sqrt{\alpha^{1/2} \omega}\,z)     
\label{osc}
\end{eqnarray}
where $ H_{N_i} $ $\;\;i=1,2,3 $ are Hermite polynomials and 
E satisfies,
\begin{equation}
E^2 \,- \,2 \, (\,N_1 \,+\,N_2 \,+\,N_3\,)\; \alpha^{1/2} \omega\,+\,
\alpha \, = \, 0
\label{Neion}
\end{equation}
with $N_i \in {\cal N}$ (natural numbers) $(i = 1,2,3)$. 

Finally we obtain for $\hat{\Psi}$

\[ \hat{\Psi}_{_{\alpha\;\,N_1\,N_2\,N_3}}  =
A_{_{\alpha\;\,N_1\,N_2\,N_3}} 
\,e^{-i\alpha \,\tau\,/\,2\,\overline{m}}\, 
e^{-i E(\alpha,N_1,N_2,N_3)\,t}\, e^{-\,(\,x^2\,+\,y^2\,+\,z^2\,)\;
\alpha^{1/2} \, \omega^2 \,/\,2 } \]
\begin{equation}
\times  H_{_{N_1}}(\sqrt{\alpha^{1/2} \omega}\,x)
\;H_{_{N_2}}(\sqrt{\alpha^{1/2} \omega}\,y)\;
H_{_{N_3}}(\sqrt{\alpha^{1/2} \omega}\,z)     
\end{equation}
with $A_{_{\alpha\;\,N_1\,N_2\,N_3}}$ the normalization constant given by
\begin{equation}
A_{_{\alpha\;\,N_1\,N_2\,N_3}}\,=\,\frac{\alpha^{1/2} \omega}{\pi}  
\frac{1}{\sqrt{2\pi \overline{m}}\; 
2^{(N_1\, +\, N_2\, +\, N_3)/2} \; (N_1 !\; N_2 ! \; N_3 !)^{1/2}}
\end{equation}

We now turn to the calculation of $\Psi$ in the case $\alpha < 0$.
The equation (\ref{16}) is now
\begin{equation}
(- \, \bigtriangleup + \,\alpha \, \omega^2 \, r^2 \, 
 + \alpha \,) \,\; \varphi \; 
 = \;E^2 \,\varphi  
\label{020}
\end{equation}
This equation has the following eight independent solutions:

\begin{equation}
\hat{\Phi}_{\alpha E_1 E_2 E_3}(x,y,z)=
\left\{
\begin{array}{rc}
x \;y \;z & \hat{\Phi}_{E_{1}} (x) \; \hat{\Phi}_{E_{2}} (y) \; \hat{\Phi}_{E_{3}} (z) \\
x \;y  & \hat{\Phi}_{E_{1}} (x) \; \hat{\Phi}_{E_{2}} (y) \; \hat{\Phi'}_{E_{3}} (z) \\
x \;z  & \hat{\Phi}_{E_{1}} (x) \; \hat{\Phi'}_{E_{2}} (y) \; \hat{\Phi}_{E_{3}} (z) \\
x  & \hat{\Phi}_{E_{1}} (x) \; \hat{\Phi'}_{E_{2}} (y) \; \hat{\Phi'}_{E_{3}} (z) \\ 
y \;z & \hat{\Phi'}_{E_{1}} (x) \; \hat{\Phi}_{E_{2}} (y) \; \hat{\Phi}_{E_{3}} (z) \\
y  & \hat{\Phi'}_{E_{1}} (x) \; \hat{\Phi}_{E_{2}} (y) \; \hat{\Phi'}_{E_{3}} (z) \\
z  &\hat{\Phi'}_{E_{1}} (x) \; \hat{\Phi'}_{E_{2}} (y) \; \hat{\Phi}_{E_{3}} (z) \\
 & \hat{\Phi'}_{E_{1}} (x) \; \hat{\Phi'}_{E'_{2}} (y) \; \hat{\Phi'}_{E_{3}} (z)  
\end{array}
\right.
\end{equation}

with
\begin{equation}
\hat{\Phi}_{E_{1} \alpha} (x) = \;\Phi \,\left(\,\frac{3}{4} \,
+ \,\frac{i\,(E_{1}^2 \,- \,\frac{\alpha}{3})}{4 \,(-\alpha)^{1/2} \, \omega} \, , \,
\frac{3}{2} \, , \, i\, \sqrt{- \, \alpha}\, \omega \, x^2 \,\right) \, 
\end{equation}

\begin{equation}
\hat{\Phi}_{E_{2} \alpha} (y) = \;\Phi \,\left(\,\frac{3}{4} \,
+ \,\frac{i\,(E_{2}^2 \,- \,\frac{\alpha}{3})}{4 \,(-\alpha)^{1/2} \, \omega} \, , \,
\frac{3}{2} \, , \, i\, \sqrt{- \, \alpha}\, \omega \, y^2 \,\right) \, 
\end{equation}

\begin{equation}
\hat{\Phi}_{E_{3} \alpha} (z) = \;\Phi \,\left(\,\frac{3}{4} \,
+ \,\frac{i\,(E_{3}^2 \,- \,\frac{\alpha}{3})}{4 \,(-\alpha)^{1/2} \, \omega} \, , \,
\frac{3}{2} \, , \, i\, \sqrt{- \, \alpha}\, \omega \, z^2 \,\right) \, 
\end{equation}

\begin{equation}
\hat{\Phi'}_{E_{1} \alpha} (x) = \;\Phi \,\left(\,\frac{1}{4} \,
+ \,\frac{i\,(E_{1}^2 \,- \,\frac{\alpha}{3})}{4 \,(-\alpha)^{1/2} \, \omega} \, , \,
\frac{1}{2} \, , \, i\, \sqrt{- \, \alpha}\, \omega \, x^2 \,\right) \, 
\end{equation}

\begin{equation}
\hat{\Phi'}_{E_{2} \alpha} (y)  = \;\Phi \,\left(\,\frac{1}{4} \,
+ \,\frac{i\,(E_{2}^2 \,- \,\frac{\alpha}{3})}{4 \,(-\alpha)^{1/2} \, \omega} \, , \,
\frac{1}{2} \, , \, i\, \sqrt{- \, \alpha}\, \omega \, y^2 \,\right) \, 
\end{equation}

\begin{equation}
\hat{\Phi'}_{E_{3} \alpha} (z)  = \;\Phi \,\left(\,\frac{1}{4} \,
+ \,\frac{i\,(E_{2}^2 \,- \,\frac{\alpha}{3})}{4 \,(-\alpha)^{1/2} \, \omega} \, , \,
\frac{1}{2} \, , \, i\, \sqrt{- \, \alpha}\, \omega \, z^2 \,\right) \, 
\end{equation}
where $\Phi (\alpha,\beta,s) $ are the degenerate hypergeometric functions 
\cite{gra} and  $E_{_1}^2 + E_{_2}^2 + E_{_3}^2 = E^2$ are real numbers.
Thus we obtain for $\hat{\Psi}$ the expression:
\begin{eqnarray}
\hat{\Psi}_{_{\,N_1\,N_2\,N_3\,\alpha}} & = & 
A_{_{N_1\,N_2\,N_3\,\alpha}} 
\,e^{-i\, \alpha \,\tau\,/\,2\,\overline{m}}\, 
e^{-\,i\,E\,t} \nonumber \\ 
 & \times & e^{-i\,(\,x^2\,+\,y^2\,+\,z^2\,)\, \sqrt{-\alpha}\, \omega/2 } \, 
\hat{\Phi}_{_{\alpha E_1 E_2 E_3}}(x,y,z)
\end{eqnarray}
with

\[ A_{_{\alpha E_1 E_2 E_3}} = \left( |E_1| |E_2| |E_3| \right) \times \] 
\begin{equation}
\frac{\left| \Gamma \left( \frac{1}{4} + \frac{i({E_1}^2 - \alpha/3)}
{4(-\alpha)^{1/2} \omega} \right) \right|
\left| \Gamma \left( \frac{1}{4} + \frac{i({E_2}^2 - \alpha/3)}
{4(-\alpha)^{1/2} \omega} \right) \right|\left| \Gamma \left( \frac{1}{4} 
+ \frac{i({E_3}^2 - \alpha/3)}{4(-\alpha)^{1/2} \omega} \right) \right| }
{2^5 \pi^{7/2} \omega^{3/2} (-\alpha)^{3/4} \overline{m}^{1/2}}
\end{equation}
The function $\hat{\Psi}$ has been normalized to
\begin{equation}
\delta(\alpha - \alpha') \delta(E_1 - E_1') \delta(E_2 - E_2') \delta(E_3 - E_3')
\end{equation}
Finally the general solution is given by

\[\Psi(x_{\mu}, \tau) = \int_{0}^{\infty} d\alpha 
\sum_{N_1 N_2 N_3} C_{_{\alpha N_1 N_2 N_3}}
e^{-\,i\,E(\alpha N_1 N_2 N_3)\,t} e^{-i\, \alpha \,\tau\,/\,2\,\overline{m}}
e^{-i\,(\,x^2\,+\,y^2\,+\,z^2\,)\, \sqrt{-\alpha}\, \omega/2 }\] 
\[\times  H_{_{N_1}}(\sqrt{\alpha^{1/2} \omega}\,x) 
H_{_{N_2}}(\sqrt{\alpha^{1/2} \omega}\,y) 
H_{_{N_3}}(\sqrt{\alpha^{1/2} \omega}\,z) \;\;+ \]
\[\int_{-\infty}^{0} d\alpha \int_{-\infty}^{+\infty} dE_1 dE_2 dE_3
C_{_{\alpha N_1 N_2 N_3}} e^{-\,i\,\sqrt{{E_1}^2 +{E_2}^2 + {E_3}^2}\,t}\]
\begin{equation}
\times e^{-i\, \alpha \,\tau\,/\,2\,\overline{m}}
e^{-i\,(\,x^2\,+\,y^2\,+\,z^2\,)\, \sqrt{-\alpha}\, \omega/2 }
\phi_{_{\alpha E_1 E_2 E_3}}(x,y,z)
\end{equation}

\section{Discussion}

\indent In this paper we have shown how to solve a model with a simple  
interaction in the proper time formalism. 
We have presented the complete set of eigenfunctions 
for the harmonic oscillator KG equation. 
In particular, we have worked out this problem in the frame of 
the Rigged Hilbert Space with Tempered Ultradistributions and their inverse 
Fourier transform space (exponentially increasing distributions).
We presented an example of finding pseudodifferential motion
equation solutions by the use of ultradistributions.

\indent From the analysis of our equations, we can say that it 
is possible to select $ a(x_{\mu},\alpha) = 0$. In this case, 
we obtain for fixed 
$ \alpha >0 $, as a particular case, the energy spectrum which
coincides with the  usual KG harmonic oscillator solution, identifying 
$m = \alpha^{1/2}$ (being $ m $ the mass of the oscillator). 
This can be deduced from the 
dispersion relation (see eq. (\ref{Neion})). The 
eigenfunctions given by eq. (\ref{osc})
are the same of those for the usual KG harmonic oscillator.
It is clear from the results of this paper that contrary to the
usual KG formalism, in the proper time model,
solutions for negative values of $\alpha$ do exist. 
These solutions represent tachyonic particles. On the contrary to the
usual quantum field theory, solutions with $ \alpha < 0 $ are
well-behaved in the sense that they are oscillating (as for
a normal bradyonic particle), instead of
exponentially increasing. 
We want to stress that the solutions we have found are valid for
arbitrary values of the parameter asociated with the mass.
 
\indent In summary, we have found the solutions for the eigenvalues and
eigenfunctions for the HO in the proper time formalism.
Our contribution is twofold. On one hand, this results contributes to
enrich the study of relativistic quantum systems in terms of the 
proper time theory.
On the other hand, our treatment for the harmonic oscillator could
be relevant to current problems in contemporary physics. 
The generalization of our problem to the case of two interacting
particles via an harmonic oscillator potential, will be very
important to describe bound states. In order to do so, 
further considerations must be taken into account as de reduction
of motion of the mass center plus a relative movement.
We hope to report about this issue in a forthcoming paper.

\section{Acknowledgements}
We are grateful to L. Epele, H. Fanchiotti and C. Garc\'{\i}a Canal
for fruitful discussions. We also
aknoweldege L.A. Anchordoqui, C. Na\'on , S. Perez Bergliaffa and D. Torres
for insightful comments and critical reading of the manuscript.
The work of M. C. Rocca was supported in part by 
Consejo Nacional de Investigaciones Cientificas, Argentina.

\end{document}